\documentclass[prb,aps]{revtex4}
\usepackage{graphicx}

\begin{document}

\title{Nonlinear Electrical Spin Conversion in a Biased Ferromagnetic Tunnel Contact}
\vspace*{3mm}
\author{R. Jansen$^1$, A. Spiesser$^1$, H. Saito$^1$, Y. Fujita$^{1,2}$, S. Yamada$^2$, K. Hamaya$^{2,3}$ and S. Yuasa$^1$}
\affiliation{$^1\,$Spintronics Research Center, National Institute of Advanced Industrial Science and Technology (AIST), Tsukuba, Ibaraki, 305-8568, Japan.\\
$^2\,$Department of Systems Innovation, Graduate School of Engineering Science, Osaka University, Toyonaka 560-8531, Japan.\\
$^3\,$Center for Spintronics Research Network, Graduate School of Engineering Science, Osaka University, Toyonaka 560-8531, Japan.}

\begin{abstract}
The conversion of spin information into electrical signals is indispensable for spintronic technologies. Spin-to-charge conversion in ferromagnetic tunnel contacts is well-described using linear (spin-)transport equations, provided that there is no applied bias, as in nonlocal spin detection. It is shown here that in a biased ferromagnetic tunnel contact, spin detection is strongly nonlinear. As a result, the spin-detection efficiency is not equal to the tunnel spin polarization. In silicon-based 4-terminal spin-transport devices, even a small bias (tens of mV) across the Fe/MgO detector contact enhances the spin-detection efficiency to values up to 140 \% (spin extraction bias) or, for spin injection bias, reduces it to almost zero, while, parenthetically, the charge current remains highly spin polarized. Calculations reveal that the nonlinearity originates from the energy dispersion of the tunnel transmission and the resulting nonuniform energy distribution of the tunnel current, offering a route to engineer spin conversion. Taking nonlinear spin detection into account is also shown to explain a multitude of peculiar and puzzling spin signals in structures with a biased detector, including two- and three-terminal devices, and provides a unified, consistent and quantitative description of spin signals in devices with a biased and unbiased detector.
\end{abstract}

\maketitle

\section{INTRODUCTION}
\indent Whereas electronic devices are based on charge, current and voltage, the key elements of spintronics technology are spin, spin current and spin accumulation\cite{zutic,fabianacta,fertnobel,sugaharanitta,jansennmatreview}. Nevertheless, the vast majority of spin-based devices and systems employ the conversion of spin information into electrical signals. This enables the detection of the spin information and provides an electrical output that can be linked to conventional electronic circuitry. It is thus indispensable to develop efficient methods for the electrical detection of spin and to obtain a thorough understanding of the process of spin conversion. Motivated by the successful application of ferromagnetic tunnel junctions\cite{moodera,parkin,yuasa} in magnetic hard-discs, sensors and magnetic random access memory\cite{chappert,mram1,mram2,mram3}, the prevalent method for the detection of spin accumulation in a nonmagnetic material employs ferromagnetic tunnel contacts. At a tunnel interface between a nonmagnetic material and a ferromagnetic tunnel contact, a spin accumulation is converted into a charge voltage across the contact\cite{fabianacta,fert,jansensstreview}, owing to spin-polarized tunneling\cite{meservey1,meservey2}. Spin detection with ferromagnetic tunnel contacts has proven to be efficient and robust, and avoids problems due to spin absorption by the ferromagnet and the conductivity mismatch\cite{fert,schmidt,rashba}.\\
\indent Examining the rich literature on the subject, we find that the degree of understanding of electrical spin conversion at a ferromagnetic tunnel interface depends critically on whether the tunnel contact is biased or not. In the so-called nonlocal spin transport devices\cite{johnson,jedema}, one employs a four-terminal measurement configuration in which one ferromagnetic contact is biased in order to induce a spin accumulation in the nonmagnetic channel, whereas the second ferromagnetic contact, the spin detector, remains unbiased. The spin signals observed in such nonlocal devices with tunnel contacts \cite{valenzuela,tombros,crowellnphys,ciorga,suzukiAPEX2011,shiraishi,toshibanonlocal,spiesserpra,hamayanonlocalge} are well described by the theory for spin injection and detection in the linear response regime\cite{fert,maekawa}. Consequently, nonlocal devices have been instrumental to prove and understand spin injection and transport in a wide variety of materials, although they are of little direct technological relevance.\\
\indent In all other devices, including the technologically relevant ones, the ferromagnetic tunnel contact in which the spin conversion occurs is located in the current path and is thus biased. For instance, in two-terminal magnetoresistance devices having two ferromagnetic contacts on a nonmagnetic channel, the current is applied between the two ferromagnetic contacts and the spin signal is obtained from the two-terminal voltage between the two contacts. Importantly, for devices in which the spin detector contact is biased, the observed spin signals are surprising and puzzling
\cite{crowellprl,crowellbiasednonlocal,sasaki2,hamayaando,hamayaando2,ishikawa3t,kameno,bruski,sasaki,shiraishipra,ishikawa3t2,shiraishitahara,crowellbiasednonlocal2,tanakasi3t,weissesaki}, and no suitable explanation for the peculiar behavior is available. Moreover, when existing (linear) transport theories are applied to devices with a biased detector, the conclusions are inconsistent with those obtained from analysis of nonlocal spin transport devices, even if the same structure is used for the different measurement configurations.\\
\indent It is shown here that spin detection in a biased ferromagnetic tunnel contact cannot be described by linear transport equations, i.e., spin detection is nonlinear. The nonlinearity originates from the nonuniform energy distribution of the tunnel current. The deviation from linear transport is surprisingly strong and already appears at small bias for which charge transport is still linear. It is also demonstrated that nonlinear spin detection is the common origin of the various puzzling and inexplicable spin signals in devices with a biased detector\cite{crowellprl,crowellbiasednonlocal,sasaki2,hamayaando,hamayaando2,ishikawa3t,kameno,bruski,sasaki,shiraishipra,ishikawa3t2,shiraishitahara,crowellbiasednonlocal2,tanakasi3t,weissesaki}. In fact, taking the nonlinearity of spin detection into account (i) explains these puzzling results qualitatively and quantitatively, and (ii) provides a unified description of electrical spin signals in devices with and without biased detector, in which the spin signals and spin transport parameters obtained from devices with a biased detector are consistent with those extracted from nonlocal spin-transport measurements.\\

\section{BACKGROUND}
{\bf Linear transport theory of spin detection.} Two important quantities define the spin transport across a tunnel contact between a ferromagnet and a nonmagnetic material (metal or semiconductor). The first is the spin polarization $P_I$ of the charge tunnel current $I$, generating a spin current $I_s=P_I\,I$. The second quantity is the spin-detection efficiency, denoted by $P_{det}$ and defined as $V_{spin}$/($\Delta\mu$/2\,e), with $V_{spin}$ the extra voltage across the tunnel contact produced by a spin accumulation $\Delta\mu$ in the nonmagnetic material ($e$ is the electron charge). In a linear transport description, due to reciprocity, these two quantities are necessarily identical ($P_I$ = $P_{det}$), which is straightforward to show. We start from the linear expressions \cite{fert,jansensstreview} for the tunnel currents for majority ($\uparrow$) and minority ($\downarrow$) spin electrons:
\begin{eqnarray}
I^{\uparrow} = G^{\uparrow}\,\left(V-\frac{\Delta\mu}{2\,e}\right) \label{eqn1}\\
I^{\downarrow} = G^{\downarrow}\,\left(V+\frac{\Delta\mu}{2\,e}\right) \label{eqn2}
\end{eqnarray}
with $G^{\uparrow}$ and $G^{\downarrow}$ the tunnel conductance for majority and minority spin electrons, respectively. Defining the tunnel resistance as $R_{tun}=1/(G^{\uparrow}+G^{\downarrow})$ and the spin polarization of the tunnel conductance as $P_G=(G^{\uparrow}-G^{\downarrow})/(G^{\uparrow}+G^{\downarrow})$,
we readily obtain $V=R_{tun}\,I + P_G\,\Delta\mu/2\,e$. The last term represents the spin voltage, from which we obtain $P_{det} = P_G$. Secondly, the spin current generated by a bias across the contact is equal to $P_G\,I$ (provided that the spin accumulation is small so that back flow (conductivity mismatch) is negligible). Hence, both the quantities $P_I$ and $P_{det}$ are equal to $P_G$. Within the linear description, the spin-detection efficiency cannot be different from the spin polarization of the charge tunnel current. This statement holds at any particular bias, i.e., even if $P_G$ depends on bias voltage, as it generally does, the linear transport equations (\ref{eqn1}) and (\ref{eqn2}) still dictate that $P_I$ and $P_{det}$ are both equal to $P_G$, which means that $P_I$ and $P_{det}$ have the same variation as a function of bias. Thus, in the terminology used here and elsewhere\cite{vanwees}, a bias-dependent $P_G$ does not constitute nonlinearity. In the following we will use a four-terminal spin transport device in various configurations to independently measure both the spin polarization of the charge current and the spin-detection efficiency. It will be shown that these two quantities are completely different when the ferromagnetic tunnel contact is under bias, implying that a linear transport description is not appropriate.\\

\section{RESULTS}
{\bf Spin polarization of current.} The spin current produced by a charge current across a ferromagnetic tunnel contact is measured using the standard nonlocal geometry (Fig 1a). The 4-terminal devices have a 70 nm thick heavily doped n-type Si channel, two Fe/MgO tunnel contacts for spin injection and detection, and two nonmagnetic (Ti/Au) reference contacts at the ends of the Si strip. When a bias current is applied to one of the Fe/MgO contacts, a spin accumulation is induced in the Si channel. Spin diffusion causes the spin accumulation to spread out through the channel, resulting in a finite spin accumulation under the second Fe/MgO contact that acts as detector, converting the spin accumulation into a charge voltage $V_{NL}$. Importantly, the charge current in the detector contact is kept at zero. The exact same devices have recently been used\cite{spiesserpra} to establish that a giant spin accumulation can be created in the Si, with the spin splitting reaching 13 meV at 10 K and 3.5 meV at room temperature. All the nonlocal spin transport data (spin-valve geometry as well as Hanle spin precession) on these devices are perfectly well described by the existing theory for spin injection and diffusion coupled with the linear response expression for spin detection in an unbiased detector. All the relevant spin transport parameters are thus known\cite{spiesserpra} (at 10 K, the spin-diffusion length $L_{SD}$ is 2.2 $\mu$m, the spin lifetime is 18 ns, and the tunnel spin polarization of the Fe/MgO contacts is 53 \% at low bias).\\

\begin{figure}[htb]
\centering
\includegraphics*[width=155mm]{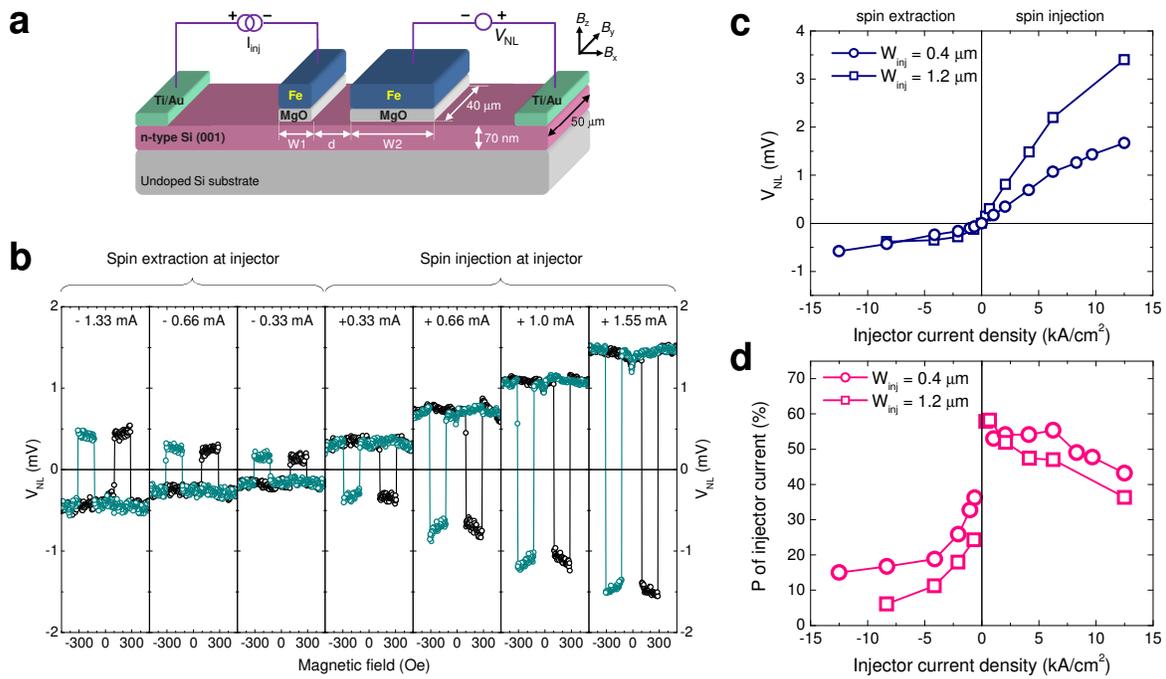}
\caption{Spin polarization of injector current across a Fe/MgO contact in a Si nonlocal spin-transport device. (a) Schematic layout of the nonlocal device having a n-type Si channel and two Fe/MgO contacts, with dimensions indicated ($W$1 = 0.4 $\mu$m, $W$2 = 1.2 $\mu$m, $d$ = 0.4 $\mu$m). The MgO is 2 nm thick, as determined before\cite{spiesserpra}. (b) Nonlocal spin signals $V_{NL}$ measured in spin-valve geometry with the external magnetic field applied in-plane (B$_Y$) along the long axis of the FM contacts, for different values of the current across the injector Fe/MgO contact. Positive current corresponds to injection of electron spins from the Fe into the Si channel. The narrow FM strip (0.4 $\mu$m) was used as injector, and the spin accumulation in the Si was detected using the wide FM contact (1.2 $\mu$m), while keeping the current across the detector strictly at zero. The B$_Y$ is either swept from plus to minus (green symbols), or in the opposite direction (black symbols). Field-independent offsets were subtracted from the measured signals. (c) Nonlocal spin-valve signals as a function of current density across the injector contact measured using either the narrow FM strip (0.4 $\mu$m) as injector and the wide FM strip (1.2 $\mu$m) as nonlocal detector (circles), or vice versa, using the wide FM strip (1.2 $\mu$m) as injector and the narrow FM strip (0.4 $\mu$m) as the nonlocal detector (square symbols). (d) Extracted spin polarization of the current across the injector Fe/MgO contact. All data was obtained at T = 10 K.}
\label{fig1}
\end{figure}

\indent The nonlocal spin-valve signals were measured for different magnitude and sign of the current across the 0.4 $\mu$m wide Fe/MgO injector contact (Fig. 1b). The 1.2 $\mu$m wide Fe/MgO contact is the nonlocal detector. The amplitude of the nonlocal spin-valve signal increases as the injector current increases and $V_{NL}$ also changes sign when $I$ does. The spin signal at positive bias (electron spin injection from the Fe into the Si) is larger than that at negative bias (electron spin extraction from the Si to the Fe), as can be seen by plotting $V_{NL}$ as a function of current density (Fig. 1c, circles). The amplitude of the nonlocal spin signal is proportional to $P_{det}\,P_I\,I$, multiplied by a factor that includes an integration over time and space containing geometric factors and spin-transport parameters that are all known. Importantly, since the detector current is kept constant (i.e., at zero), $P_{det}$ is constant. Therefore, the variation of the spin signal is completely due to the variation of the spin current $P_I\,I$ across the 0.4 $\mu$m wide injector contact. Since $I$ is known, we can extract $P_I$ from the data for each injector bias (Fig. 1d, circles). The current polarization has a value of around 53 \% at low bias, decreases with increasing current density, particularly for spin extraction (negative bias). A similar set of spin-valve curves was obtained (not shown) with the injector and detector contact interchanged, thus using the 1.2 $\mu$m wide Fe/MgO contact as injector, and the 0.4 $\mu$m wide Fe/MgO contact as the nonlocal detector. From this data one obtains the spin polarization of the charge current across the 1.2 $\mu$m contact (square symbols in Fig. 1c and d). The behavior of the two contacts is similar, although not exactly the same.\\
\indent The decay of $P_I$ at larger injector bias, and the asymmetry with respect to the current polarity, are consistent with previous reports \cite{valenzuela,park}. The decay is because the tunnel spin polarization of the electrons tunneling out of or into the Fe decreases away from the Fermi energy of the Fe. The $P_I$ decays faster for negative bias, because it involves electrons tunneling into empty states in the Fe well-above the Fermi energy of the Fe. The decay is relatively weak for positive bias, because in this case a large fraction of the tunneling electrons still originates from the Fermi energy of the Fe or just below it \cite{valenzuela,park}.\\
\\
{\bf Spin-detection efficiency.} Next, we use the same device to determine the spin-detection efficiency. The current across the 0.4 $\mu$m wide Fe/MgO contact (the injector) is fixed at a value of $+$1 mA. The other Fe/MgO contact is used as nonlocal spin detector, but a nonzero bias current $I_{det}$ is applied across it using a separate current supply (Fig. 2a). Because the differential equations for spin-diffusion in a non-magnetic material are linear in the spin accumulation, the spin accumulations from different sources (injector and detector) are simply superimposed. Moreover, as pointed out before\cite{crowellbiasednonlocal2}, in such a nonlocal measurement with a biased detector, the nonlocal spin-valve signal is not affected by the extra spin accumulation $\Delta\mu_{det}$ that is induced in the channel by the detector current, because it produces a spin voltage that is proportional to $P_{det}\,(\Delta\mu_{det}/2\,e)$ at the detector interface and that does not depend on the relative magnetization alignment of injector and detector. Hence, the spin-valve signal is still given by $P_{det}\,(\Delta\mu_{inj}/2\,e)$, as in the conventional nonlocal measurement, but now the spin accumulation $\Delta\mu_{inj}$ induced by the injector current is kept constant and $P_{det}$ changes as a function of the bias across the detector.\\

\begin{figure}[htb]
\centering
\includegraphics*[width=155mm]{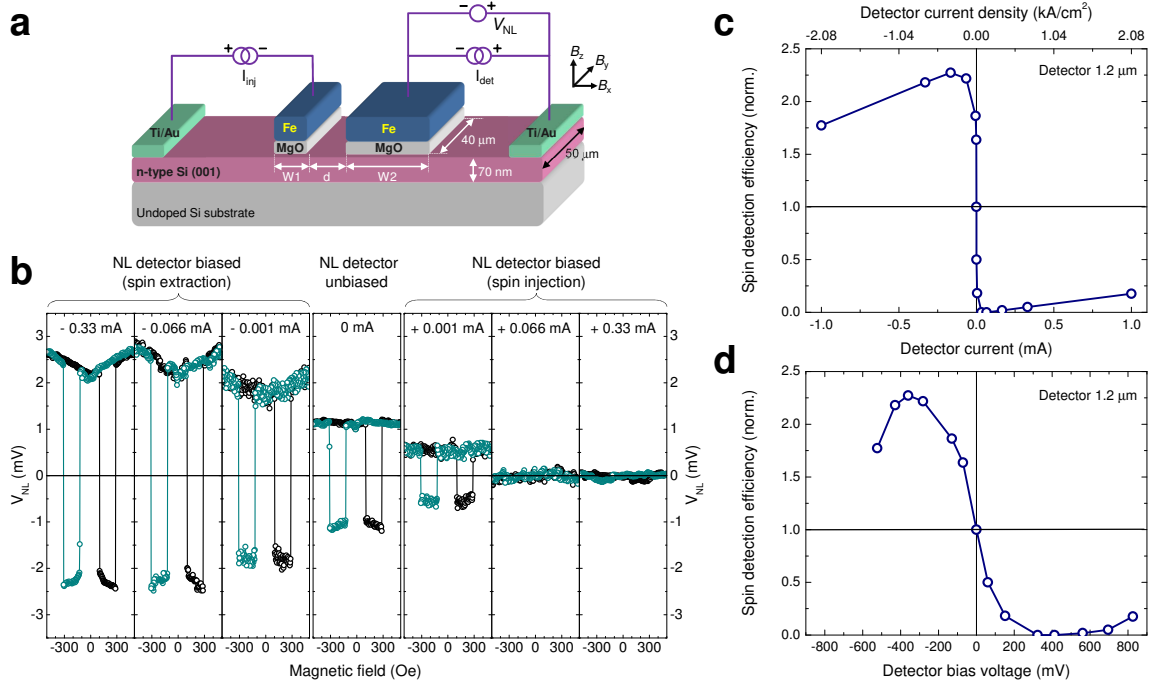}
\caption{Spin-detection efficiency of a biased Fe/MgO detector contact in a Si nonlocal spin-transport device. (a) Schematic layout of the nonlocal device with a biased detector. (b) Nonlocal spin signals $V_{NL}$ measured in spin-valve geometry for a fixed value ($+$1 mA) of the current across the injector, but different values of the current across the detector contact, as indicated. Positive detector current corresponds to electrons flowing from the Fe/MgO detector to the Si. The narrow FM strip (0.4 $\mu$m) was used as injector, and the wide FM contact (1.2 $\mu$m) was the biased detector, thus measuring the spin-detection efficiency of the wide contact. The B$_Y$ is either swept from plus to minus (green symbols), or in the opposite direction (black symbols). Non-trivial \cite{invertedhanle} background signals due to the nonzero current in the detector were experimentally determined by measuring the nonlocal voltage across the biased detector with the {\em injector} current switched off and subtracted from the data (Supplemental Material \cite{supp}). Spin-valve data for larger detector bias is given in the Supplemental Material \cite{supp}. (c),(d) Extracted spin-detection efficiency as a function of current (c) or the bias voltage (d) across the detector. The spin-detection efficiency is normalized to the value of an unbiased detector using the $V_{NL}$ signal for 0 mA detector current (central panel of (b)). T = 10 K.}
\label{fig2}
\end{figure}

\indent The spin-valve signals (Fig. 2b) are rapidly reduced for positive bias across the detector contact\cite{notePdet}. A significant reduction already occurs at a very small current of only $+$0.001 mA, corresponding to a bias of only $+$60 mV. The spin signal completely disappears for $+$0.066 mA. For the negative bias, the effect is strong as well, but opposite, i.e., the spin signal with bias across the detector is larger than for the unbiased detector. The enhancement is already significant at a very small current of $-$0.001 mA (bias of $-$71 mV), and an enhancement by more than a factor of two is achieved for larger negative bias. From this data, we extract the spin-detection efficiency of the biased detector tunnel contact relative to the spin-detection efficiency at zero bias (Figs. 2c and 2d). We observe that the spin-detection efficiency is changed drastically by a bias across the detector, and that pronounced changes occur already at small bias (below $\pm$200 mV).\\
\indent Having determined the spin polarization of the current (Fig. 1) and the spin-detection efficiency (Fig. 2), we next compare the two quantities (see Fig. 3). The key result is that, when the Fe/MgO tunnel contact on Si is biased, the spin-detection efficiency is completely different from the spin polarization of the current. Whereas the $P_I$ decays when a bias is applied and most strongly so for negative bias, the spin-detection efficiency is enhanced and largest for negative bias (spin extraction), and is reduced rapidly to zero for positive bias (spin injection), for which $P_I$ remains rather large. These results are inconsistent with what is expected from linear transport theory (see section I), for which the current spin polarization and the spin-detection efficiency are identical and both given by the tunnel spin polarization $P_G$ at every bias. Also note that because at zero bias $P_{det}$ = 58 \%, the enhancement by a factor of 2.3 at negative bias implies that $P_{det}$ = 130 \% and thus larger than 100 \%. This is impossible in a linear transport description (for which $P_{det} = P_G$, and $P_G \leq 100 \%$, by definition).

\begin{figure}[htb]
\centering
\includegraphics*[width=80mm]{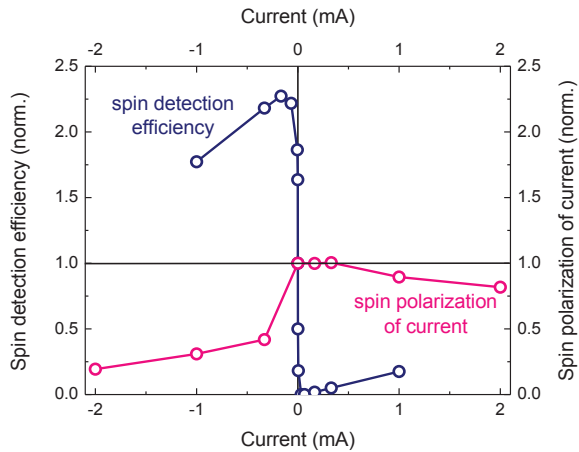}
\caption{Comparison of the spin polarization of the current and the spin-detection efficiency of a biased Fe/MgO tunnel contact on Si. The dark blue signals represent the normalized spin-detection efficiency of the 1.2 $\mu$m wide contact versus bias current (left and bottom axis), as obtained from the data in Fig. 2c. The pink signals represent the current spin polarization of the same 1.2 $\mu$m wide contact versus bias current (right and top axis), as obtained from the data in Fig. 1d (square symbols) after normalizing to the value ($\sim$ 58 \%) that the current polarization approaches at zero current. In both cases, positive current corresponds to electrons flowing from the Fe to the Si. T = 10 K.}
\label{fig3}
\end{figure}

{\bf Is spin drift responsible?} In principle, the electric field $E$ that the applied current produces in the Si channel can affect the spin signals. In previous work on Fe/GaAs devices\cite{crowellbiasednonlocal}, changes in the amplitude of the nonlocal spin-valve signal by a bias across the nonlocal detector were attributed to the electric field in the GaAs. However, based on new measurements at lower currents this interpretation was later abandoned\cite{crowellbiasednonlocal2}. Our results corroborate the latter. We observe significant changes in the nonlocal spin-valve signal at very small current ($\pm$ 0.001 mA). Using the cross section of the Si channel (width 50 $\mu$m, thickness 70 nm) and the Si resistivity (1.3 $m\Omega$cm), the corresponding electric field\cite{noteE} in the Si is only 4 V/m. This is extremely small and has no noticeable effect on the spin transport. The effect of spin drift \cite{flatte} becomes relevant above a critical electric field $E_{crit} = \varepsilon_{drift}/e\,L_{SD}$, where $\varepsilon_{drift}$ is an energy scale that, loosely speaking, is the Fermi energy for degenerate semiconductors, but more precisely, it is given by $\varepsilon_{drift}=e\,D/\mu_e$. With a diffusion constant $D$ of a few cm$^2$/s and an electron mobility $\mu_e$ of $\sim$ 200 cm$^2$/Vs, the $\varepsilon_{drift}$ is of the order of 10 meV. We then obtain $E_{crit} \approx$ 4500 V/m. The electric field of 4 V/m in our devices is thus completely negligible. Spin drift would become noticeable only for electric fields that are more than 3 orders of magnitude larger. The observed variation of the spin-detection efficiency is thus not due to spin drift.\\
\\
\\
{\bf Origin of nonlinear spin detection.} The results are explained by the nonlinearity of spin detection via tunneling. The nonlinearity is found to be inherent to tunneling and ultimately arises from the dependence of the tunnel probability on the electron energy $\varepsilon$. When a bias $V$ is applied across a tunnel contact, electrons with energy between 0 and $e\,V$ can contribute to the tunnel current, but the energy distribution of the tunnel current is peaked near the highest energies within this energy interval, because electrons with higher energy typically have a larger tunnel probability. Since the spin accumulation (in thermal equilibrium) modifies the occupation of the states very close the Fermi level in the nonmagnetic electrode $\varepsilon_{F}^{NM}$, the alignment of $\varepsilon_{F}^{NM}$ relative to the peak in the tunnel current energy distribution has a profound effect on the spin-detection efficiency.\\
\indent To illustrate and quantify this, we employ a free-electron description \cite{wolf} of tunneling using a rectangular tunnel barrier (width $w$ and height $\Phi$) sandwiched between two metallic electrodes, one of which is a ferromagnet and the other is not. The energy-resolved tunnel current is computed for each spin, using the tunnel transmission function $T(\varepsilon)$ and the Fermi-Dirac distributions of the electrons in the electrodes, which are spin split by an amount $\Delta\mu$ in the nonmagnetic electrode (details are given in Supplemental Material \cite{supp}). The tunnel spin polarization $P_G$, which in general depends on energy as well, is set to a constant value, so as to remove any effect on the variation of the spin-detection efficiency with bias.

\begin{figure}[htb]
\centering
\includegraphics*[width=150mm]{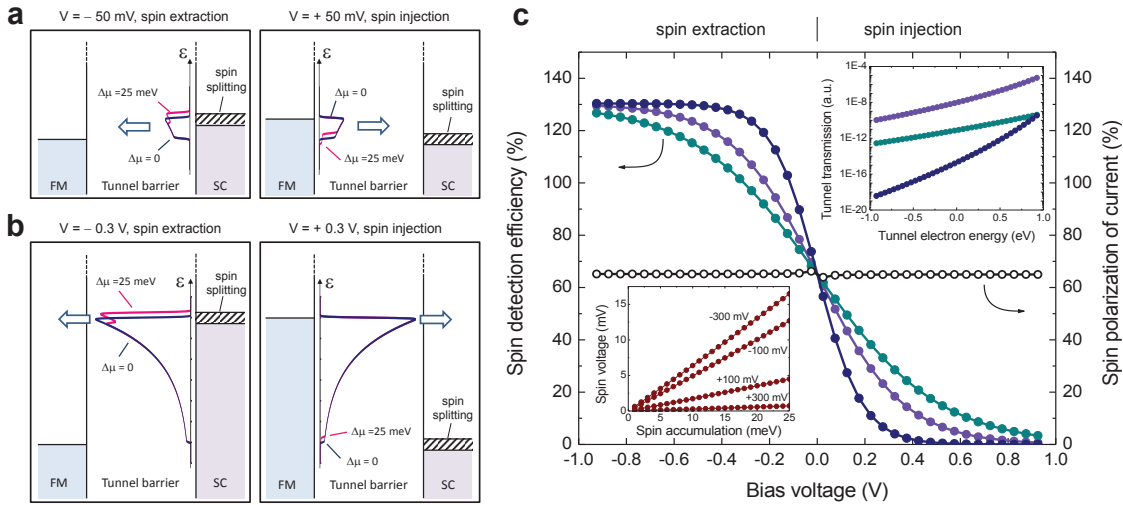}
\caption{Origin and calculation of the nonlinear spin detection. (a),(b) Schematic energy diagrams of a tunnel junction between a ferromagnet (FM) and a nonmagnetic material (SC) under small bias ($\pm$ 50 mV, (a)) or moderate bias ($\pm$ 0.3 V, (b)), with superimposed on it the distribution of the tunnel current as a function of electron energy $\varepsilon$, for two cases: (i) no spin accumulation in the SC (dark blue curves), (ii) with a spin accumulation of 25 meV in the SC (pink curves). The spin splitting in the SC is indicated by the black shaded area. Arrows indicate the direction of electron flow. It is stressed that the tunnel distributions are not schematic drawings but the result of an actual calculation using a free electron tunneling model, for T = 10 K. (c) Calculated spin-detection efficiency versus tunnel bias voltage (solid symbols, left axis) for different values of the width ($w$) and height ($\Phi$) of the tunnel barrier (dark blue: $w$ = 2.6 nm, $\Phi$ = 1.5 eV, violet: $w$ = 1.4 nm, $\Phi$ = 1.5 eV, green: $w$ = 1.4 nm, $\Phi$ = 3 eV). The simultaneously calculated spin polarization of the charge current across the tunnel contact is also shown (open symbols, right axis). The $P_G$ was set to an arbitrary\cite{note65} constant value of 65 \% for all electron energies. The right inset shows the tunnel transmission versus the electron energy for the same combinations of $w$ and $\Phi$. The left inset shows the spin voltage as a function of $\Delta\mu$ at different values of the bias (for $w$ = 2.6 nm, $\Phi$ = 1.5 eV).}
\label{fig4}
\end{figure}

\indent The calculated energy distributions of the (spin-integrated) tunnel current are depicted in Fig. 4a and b, overlayed on the schematic energy diagram of the tunnel junction for positive and negative biases of 50 mV and 0.3 V. The profiles in the absence of a spin accumulation (dark blue lines) as well as for a spin accumulation of 25 meV (pink lines) are shown. For small bias, the tunnel current is still relatively evenly distributed in energy, but at larger bias, it becomes more and more peaked near the higher energies. For a bias of 0.3 V, the effect of the spin accumulation on the tunnel current depends drastically on bias polarity. For negative bias (spin extraction), the states whose occupation is changed by the spin accumulation have energies that match with the peak in the tunnel current energy distribution. Hence, the spin accumulation produces large changes in the tunnel current and the spin-detection efficiency is maximum. In contrast, for $+$ 0.3 V (spin injection), the spin accumulation concerns states that are in the tail of the current distribution and contribute very little to the tunnel current. In this case the spin accumulation has almost no effect on the total tunnel current and the spin-detection efficiency is minimal. At smaller bias, the energy interval is smaller and hence the energy distribution is more homogeneous. Consequently, the spin-detection efficiency depends less on bias polarity, although a difference is still visible at $\pm$ 50 mV.\\
\indent The computed spin voltage, given by $V(\Delta\mu) - V(\Delta\mu = 0)$ at constant current, scales with the magnitude of $\Delta\mu$, but the slope (which equals the spin-detection efficiency) depends sensitively on the applied bias voltage (left inset of Fig. 4c). The calculated spin-detection efficiency (Fig. 4c) decays rapidly for small positive bias, and reaches a value near zero around a few hundred mV, whereas for negative bias, the spin-detection efficiency is enhanced compared to the value at zero bias. The spin-detection efficiency computed at zero bias is equal to $P_G$, the linear response result, as it should be. The simultaneously computed spin polarization of the tunnel current is $P_G$, independent of bias, and thus matches with the spin-detection efficiency only around zero bias. Calculations for different width and height of the tunnel barrier show that the nonlinearity becomes stronger when the tunnel transmission depends more steeply on energy (Fig. 4c, right inset), because this makes the tunnel distribution more strongly peaked. Thus, the calculations demonstrate that indeed the spin polarization of the tunnel current and the spin-detection efficiency are unequal under bias, and that the deviation from linear response starts at very small bias.\\
\indent Although our model calculations capture and illustrate the basic physics and describe the experimental trends, more sophisticated calculations should include the depletion region that is present in the Si at the interface with the Fe/MgO contacts, as well as the specific density of states of the semiconductor, which includes a band gap and an impurity band for heavily-doped material. Moreover, the (complex) band structure of the MgO tunnel barrier needs to be taken into account\cite{yuasareview}. Nonetheless, the described nonlinearity is a general feature of tunneling. It not only applies to tunneling through an oxide barrier, but, for instance, also to Schottky tunnel barriers and Esaki diodes. In fact, in a Schottky tunnel contact, the tunnel energy distribution is expected to be even more peaked because there is an extra increase of the tunnel probability at higher energy because the Schottky barrier is more narrow. Another point is that the calculations were done for a constant value of $P_G$. In reality, the tunneling spin polarization is known to decay at larger bias \cite{valenzuela,park}. This explains why the measured spin-detection efficiency decays at larger negative bias (Fig. 2d). However, the same effect cannot account for the upturn in the measured spin-detection efficiency at larger positive bias, which at present is not understood.\\
\indent The essence of the mechanism we described has been suggested before\cite{crowellprl}, albeit briefly. However, no quantitative analysis was made to support it, and there was no discussion about nonlinearity, the crucial role of the tunnel energy distribution, nor about enhancement of the spin detection sensitivity compared to zero bias. Moreover, the suggestion was made\cite{crowellprl} in connection with {\it three-terminal} Hanle measurements. In contrast, in experiments on nonlocal spin transport devices with a biased nonlocal detector\cite{crowellbiasednonlocal,crowellbiasednonlocal2,weissesaki}, the mechanism we describe was not considered, but instead the results were attributed to electric fields and spin drift. We argue here that the nonlinearity of spin detection is the common origin of the puzzling data obtained in all the devices in which the spin detector contact is biased, including nonlocal spin detection with a biased detector, three-terminal Hanle spin detection as well as two-terminal magnetoresistance. We demonstrate this, starting with the two-terminal magnetoresistance in the next section.\\
\\
\\
{\bf Nonlinear spin detection \& two-terminal magnetoresistance.} In devices having two nominally identical ferromagnetic tunnel contacts on a semiconductor, one would expect that the two contacts contribute equally to the two-terminal magnetoresistance, each producing half of the total spin signal. However, experiments give an entirely different result. In Si devices with Fe/MgO tunnel contacts\cite{sasaki,shiraishipra,shiraishitahara}, it was found that the complete two-terminal spin signal is generated at the interface of one of the two tunnel contacts, whereas no spin signal is detected at the other contact. An unequal contribution of the two contacts was also reported for GaAs devices with Esaki diode contacts\cite{weissesaki}. It has also been reported\cite{sasaki2,bruski,sasaki,weissesaki} that the two-terminal magnetoresistance is larger than what is expected from linear response theory using the spin-transport parameters extracted from nonlocal devices. We shall show that these puzzling results can be understood if the nonlinear spin detection is taken into account.\\
\indent We use the same devices as before, but now the two-terminal local magnetoresistance (2T-MR) is probed (Fig. 5a). In this case, the current is applied between the two ferromagnetic strips, that will hereafter be referred to as "source" and "drain" (superscripts "S" and "D", respectively). The total two-terminal spin-valve signal is about 0.4 mV at room temperature for a current of -1.2 mA (Fig. 5c, dark blue symbols). Parenthetically, the signal measured across the source contact is also 0.4 mV (Fig. 5d, dark blue symbols), whereas no spin-valve signal is detected at the drain contact (Fig. 5e, dark blue symbols). Hence, the complete spin-valve signal is produced by only one of the two contacts. This peculiar result has previously been reported for devices with heavily-doped Si \cite{sasaki} and nondegenerate Si\cite{shiraishipra,shiraishitahara}, but no suitable explanation exist. Because spin drift has been suggested\cite{sasaki,shiraishipra,shiraishitahara}, we first prove experimentally that electric fields and spin drift are not responsible\cite{noteE330rt}. For this, we introduce a modified measurement configuration, referred to as a two-terminal {\em nonlocal} geometry (Fig. 5b). The voltage measurement is the same as in the local 2T-MR geometry, but instead of applying the current between the two FM contacts, we use two separate current supplies for the source and the drain contacts, making use of the two nonmagnetic reference contacts at the far ends of the channel. The currents $I_S$ and $I_D$ are chosen such that for both Fe/MgO/Si contacts, the direction of electron flow and the tunnel current magnitude are the same as in the regular two-terminal measurement. Thus, only the charge flow and the electric field in the Si channel are changed. Therefore, if spin transport is governed exclusively by spin diffusion, the spin accumulation profile in the channel and the spin signals will be the same as in the regular local 2T-MR measurement. We find that the spin signals for the two-terminal nonlocal geometry (pink symbols in Figs. 5c,d,e) are comparable to those obtained for the local 2T-MR geometry. This proves that spin drift is negligible and is not responsible for the absence of a spin signal at the drain contact.\\
\indent The 2T-MR results can be explained if the nonlinear spin-detection efficiency is taken into account. For quantitative analysis, we performed measurements in the 2T nonlocal geometry and at 10 K, at which all the relevant parameters are known from the data in Figs. 1 and 2. The 2T nonlocal data in Fig. 6 provides one more salient feature, namely, that the contact at which the spin-valve signal is observed, depends on the polarity of the current. The complete spin-valve signal originates from the contact at which the electrons flow from the Si to the Fe (spin extraction, contact D for $+$0.33 mA, contact S for $-$0.33 mA), whereas no spin signal is detected at the contact at which the electrons flow from the Fe to the Si (spin injection, contact S for $+$0.33 mA, contact D for $-$0.33 mA).

\begin{figure}[htb]
\centering
\hspace*{7mm}\includegraphics*[width=129mm]{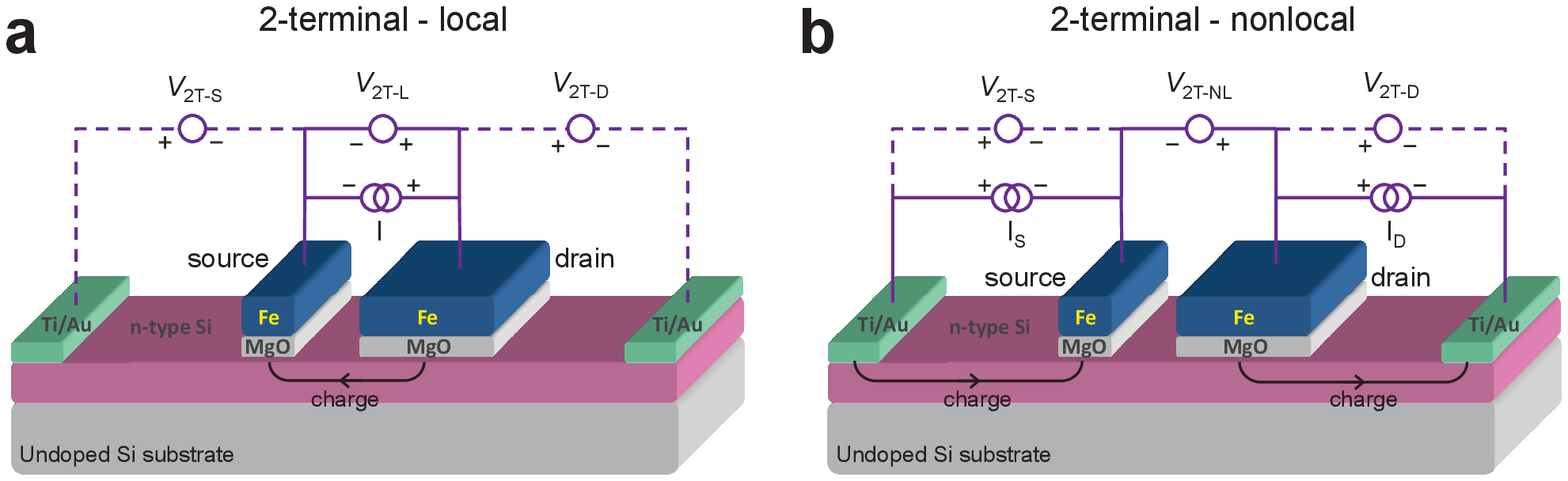}\\
\includegraphics*[width=158mm]{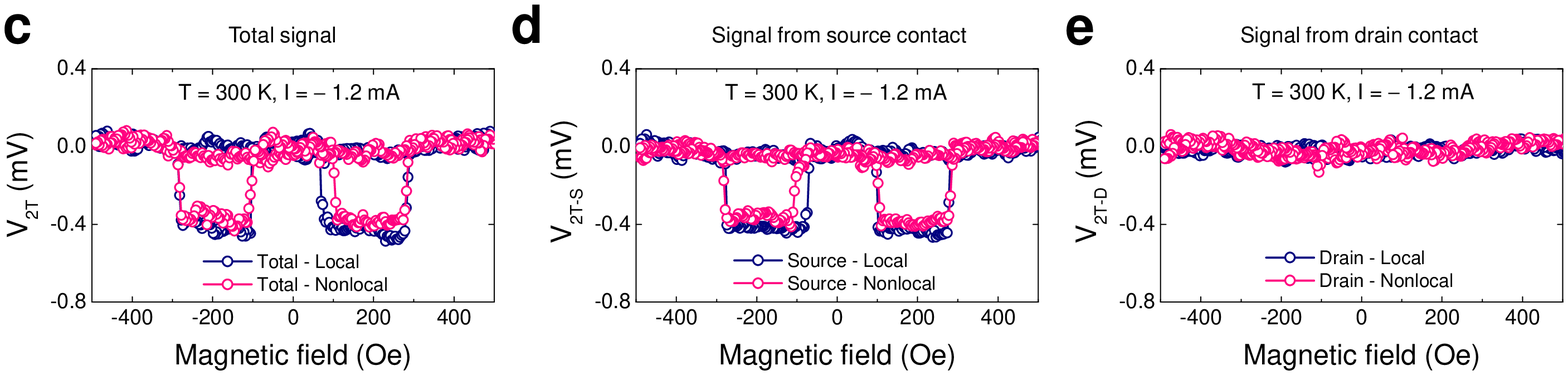}
\caption{Two-terminal MR in local and nonlocal geometry at 300 K. (a) Geometry for the measurement of the conventional local two-terminal MR, with the charge flowing between the two Fe/MgO contacts. (b) Geometry for the measurement of the two-terminal magnetoresistance in the nonlocal geometry, using a separate current supply for each of the two Fe/MgO contacts. For both geometries, the two-terminal voltage is detected in the same way, i.e., between the two Fe contacts. The narrow (0.4 $\mu$m) contact on the left is labeled as "source, S", whereas the wide (1.2 $\mu$m) contact on the right is labeled as "drain, D". (c),(d),(e) Two-terminal spin-valve signals at 300 K in local (dark blue symbols) and nonlocal (pink symbols) mode, including the total spin signal (c), and the contribution to the spin signal produced by each Fe/MgO contact separately ((d) source contact, (e) drain contact). For the local geometry, the charge current was $-$1.2 mA, corresponding to spin extraction at the source contact and spin injection at the drain contact. For the nonlocal geometry, both current supplies were set to $-$1.2 mA, which, given the wiring as indicated in (b), ensures that the direction of electron tunneling across the Fe/MgO interfaces is the same as for the local geometry. Non-trivial background signals are subtracted from the data (Supplemental Material \cite{supp}).}
\label{fig5}
\end{figure}

\begin{figure}[htb]
\centering
\hspace*{3mm}\includegraphics*[width=150mm]{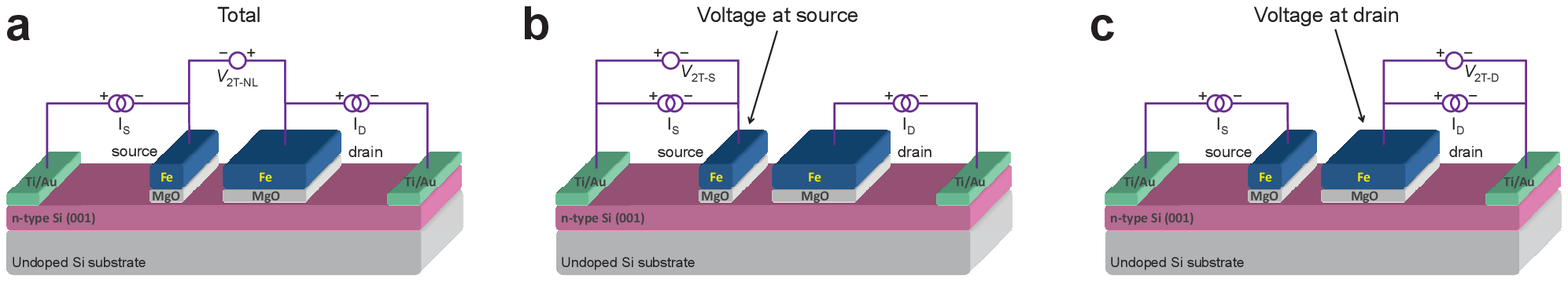}\\
\includegraphics*[width=150mm]{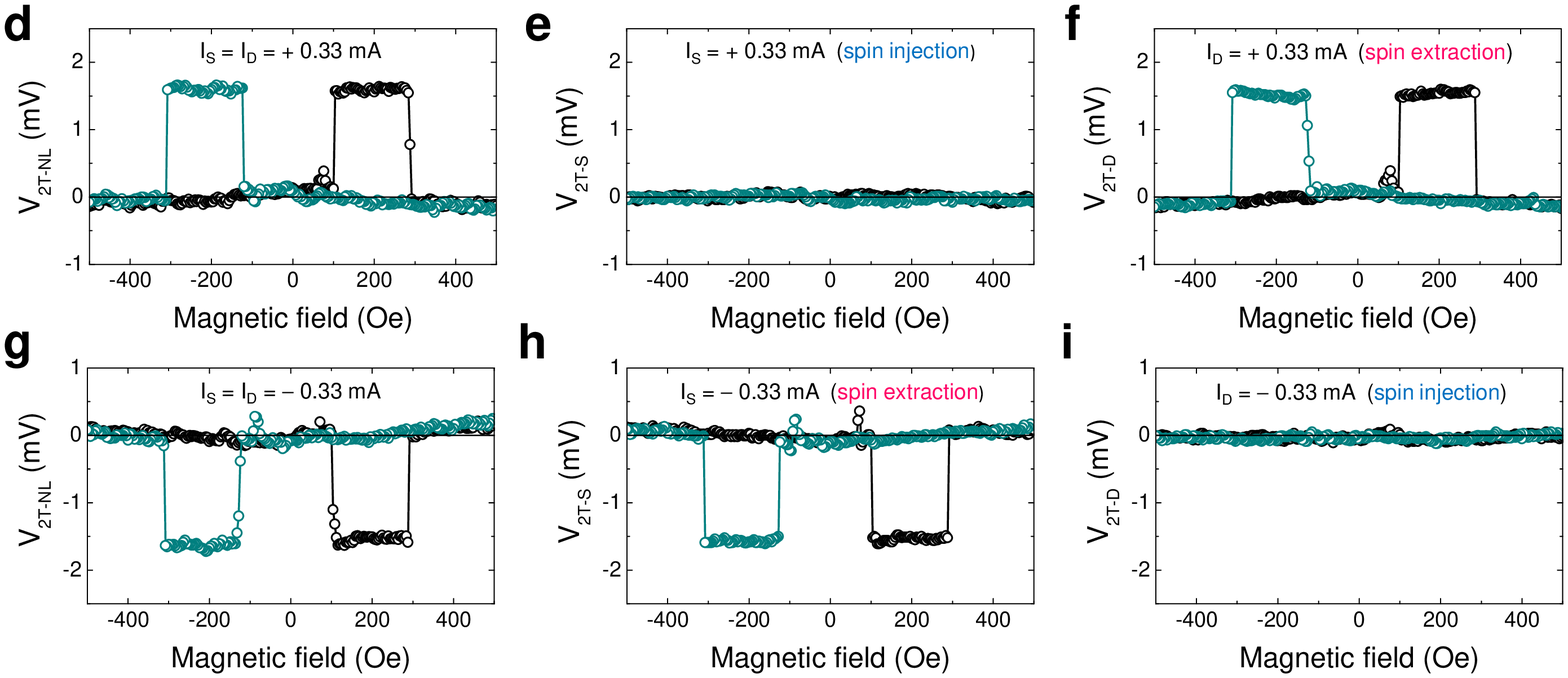}
\caption{Nonlinear spin detection in two-terminal MR at low temperature. (a),(b),(c) Geometry for the measurement of the two-terminal spin-valve signals
in nonlocal configuration. (d),(e),(f) Two-terminal spin-valve signals at 10 K in nonlocal geometry for a current of $+$0.33 mA, which corresponds to spin injection at the source contact and spin extraction at the drain contact. (g),(h),(i) Two-terminal spin-valve signals at 10 K in nonlocal geometry for a current of $-$0.33 mA, which corresponds to spin extraction at the source contact and spin injection at the drain contact. Panels (a),(d),(g) correspond to the total two-terminal signal, panels (b),(e),(h) correspond to the spin signal produced at the source contact, whereas panels (c),(f),(i) correspond to the spin signal produced at the drain contact. Background signals are subtracted from the data (Supplemental Material \cite{supp}).}
\label{fig6}
\end{figure}

\indent The experimental results are compared with the spin signals expected using, respectively, linear transport or the nonlinear detection efficiency (Table I). The total two-terminal spin-valve signal is the sum of two contributions, (i) the current across the source Fe/MgO/Si contact produces a spin accumulation in the channel with an amplitude $\Delta\mu^{S}$ under the drain contact, which converts it into a spin voltage given by $P_{det}^{D}\,(\Delta\mu^{S}/2\,e)$, and (ii) the current across the drain Si/MgO/Fe contact produces a spin accumulation in the channel with an amplitude $\Delta\mu^{D}$ under the source contact, which converts it into a spin voltage equal to $P_{det}^{S}\,(\Delta\mu^{D}/2\,e)$. These spin voltages can be computed by taking the spin signal for the conventional nonlocal measurement (Fig. 1), for which the spin detection contact is unbiased, and multiply this by the ratio $P_{det}(V)/P_{det}(0)$ of the spin-detection efficiency under bias and at zero bias. The latter ratio can be calculated in two different ways. In the first one, we use the linear transport equations, so that $P_{det}(V)=P_I(V)$ for each bias and the ratio becomes $P_{I}(V)/P_{I}(0)$ (see Table I, column labeled as "polarization ratio", with the numbers obtained from Fig. 1d). In the second case, we take the nonlinear spin detection into account. Then, the ratio $P_{det}(V)/P_{det}(0)$ is the spin-detection efficiency normalized by its value at zero bias (as previously plotted in Fig 2c,d).\\
\indent Based on the linear transport theory, one expects that the spin signals at the source and drain contact are almost identical (for instance, 0.25 mV versus 0.29 mV for $+$0.33 mA). This disagrees with the experiment, in which a spin signal is present only for one of the two contacts (Table I, last column, data from Fig. 6). Moreover, for linear transport, also the total spin signal does not agree with the experimental result. For instance, for $+$0.33 mA, the experimental signal is 1.6 mV, whereas 0.54 mV is expected. In contrast, when the nonlinear detection efficiency is taken into account, the predicted signal for each of the two contacts, as well as the predicted total spin signal, are in excellent agreement with the data. The absence of a spin signal in the contact that is under spin injection bias is because for that bias polarity, the spin-detection efficiency is negligibly small. And a large spin signal is produced at the contact at which spin extraction occurs, because for that bias polarity the spin-detection efficiency is large. This demonstrates that the peculiar behavior of the two-terminal spin signals originates from the nonlinearity of the spin detection. Taking this into account produces a qualitative as well as quantitative description of the two-terminal magnetoresistance\cite{notenondegenerate}.

\begin{table}[h]
\caption{Comparison of observed two-terminal spin signals at T = 10 K (from Fig. 6) with the values predicted, respectively, by linear transport theory and by using the nonlinear spin detection sensitivity. Predicted values of the spin signal for each contact are obtained by taking the spin signal detected with an unbiased detector (standard nonlocal geometry) and multiplying that with a factor correcting for the different detection efficiency under bias. The correction factor is either the polarization ratio (linear transport, assuming that P$_{det}$ = P$_{I}$, with P$_{I}$ obtained from Fig. 1d) or the measured relative spin-detection efficiency (obtained from figure 2c). The absolute values of the spin-detection efficiency are also indicated in brackets.} \label{table} \vspace*{2mm}
\begin{tabular}{ c c | c c c | c c c | c c c c c | c c c c c | c c c}
\hline \hline
Contact & & & Bias & & & Spin signal & & & Polarization & & Predicted signal & & & Detection & & Predicted signal & & & Observed & \\
       & & & condition & & & unbiased & & & ratio $P_{I}(V)/P_{I}(0)$ & & linear transport & & & efficiency & & nonlinear detection & & & signal & \\
\hline
 & & & $+$0.33 mA & & & & & & & & & & & & & & & & & \\
Source & & & injection & & & 0.25 mV & & & 54.1/53 & & 0.25 mV & & & 0.05 (2.7 \%) & & 0.01 mV & & & 0 mV & \\
Drain & & & extraction & & & 0.69 mV & & & 24.3/58 & & 0.29 mV & & & 2.18 (126 \%) & & 1.51 mV & & & 1.6 mV & \\
Total & & & & & & & & & & & 0.54 mV & & & & & 1.52 mV & & & 1.6 mV & \\
\hline
 & & & $-$0.33 mA & & & & & & & & & & & & & & & & & \\
Source & & & extraction & & & 0.60 mV & & & 25.9/53 & & 0.29 mV & & & 2.62 (139 \%) & & 1.57 mV & & & 1.6 mV & \\
Drain & & & injection & & & 0.33 mV & & & 58.3/58 & & 0.33 mV & & & 0.05 (2.9 \%) & & 0.02 mV & & & 0 mV & \\
Total & & & & & & & & & & & 0.62 mV & & & & & 1.59 mV & & & 1.6 mV & \\
\hline \hline
\end{tabular}
\end{table}

{\bf Nonlinear spin detection \& three-terminal spin signals.} Finally, we show that the nonlinear spin detection also explains the puzzling dependence of three-terminal spin signals on bias polarity. This issue was first noted for Fe/GaAs structures. Experiments using either optical spin detection methods or nonlocal spin transport devices have clearly established that a spin accumulation can be electrically induced in GaAs by injecting electron spins from an Fe Schottky tunnel contact into the GaAs, or by extracting spins from the GaAs into the Fe \cite{crowellnphys,crowellscience}. And yet, when in the same Fe/GaAs structures the spin accumulation under the (biased) injector contact is probed in a three-terminal geometry (using the Hanle effect), one finds that the spin accumulation produces a three-terminal Hanle signal only for spin extraction bias, but no spin signals are observed for the bias that corresponds to spin injection\cite{crowellprl}. The latter would lead one to conclude that no spin accumulation is created in the GaAs under spin injection bias, in direct contradiction with the results of optical and nonlocal detection. Similarly puzzling behavior was observed for devices with a Si channel and Schottky contacts\cite{hamayaando,hamayaando2} or MgO tunnel contacts with CoFe, Fe or Co$_2$FeSi electrodes\cite{ishikawa3t,kameno,ishikawa3t2,tanakasi3t}.\\
\indent We obtain similar results on our Si devices with Fe/MgO contacts. As an example, we present data using a 1.2 $\mu$m wide ferromagnetic contact (Fig. 7). For positive bias, no three-terminal Hanle spin signal could be detected above the noise level. However, for spin extraction bias, a clear Hanle spin signal is observed, and the line width is in agreement with what is expected\cite{notehanle}. When the same 1.2 $\mu$m wide tunnel contact is used as the injector in a 4-terminal configuration, a nonlocal spin signal is detected for both polarities (V$_{NL}$ = $+$0.3 mV for $+$0.33 mA and V$_{NL}$ = $-$0.125 mV for $-$0.33 mA, see Fig. 1b, square symbols). Hence, a spin accumulation is present in the Si channel for both polarities of the current, and the spin accumulation is larger for positive bias, in apparent contradiction with the three-terminal data. Note that the data was taken at sufficiently small current, so as to ensure that the electric field in the Si channel is negligibly small and spin drift does not play any role.

\begin{figure}[htb]
\centering
\hspace*{4mm}\includegraphics*[width=65mm]{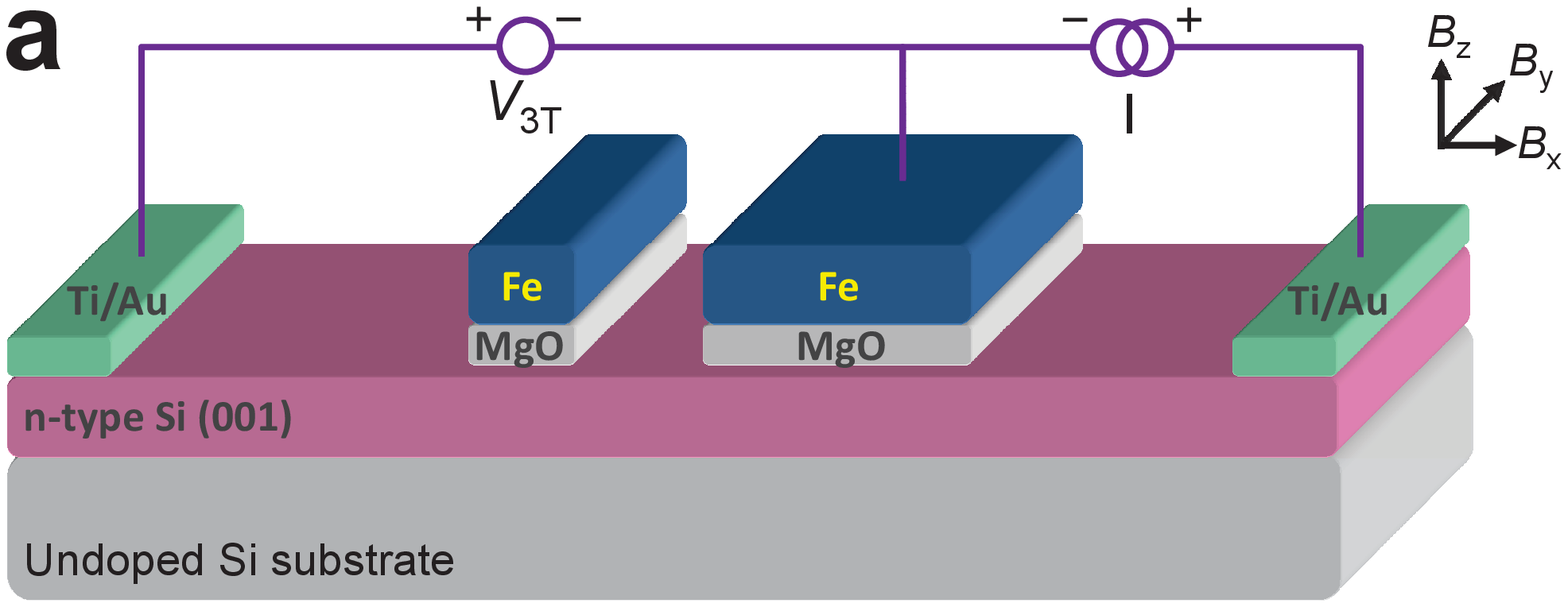}\\
\includegraphics*[width=88mm]{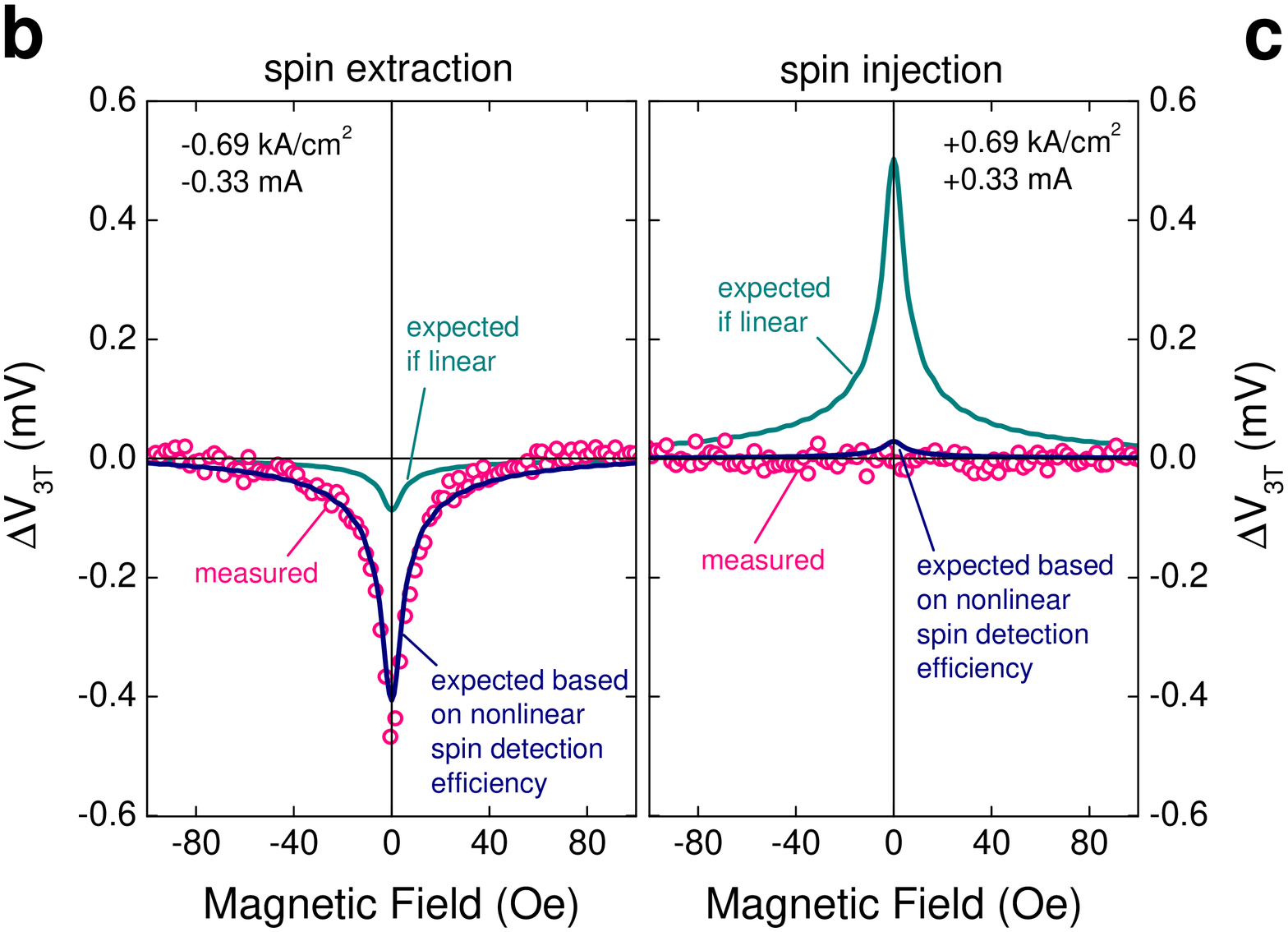}
\caption{Nonlinear spin detection in three-terminal Hanle spin signals. (a) Geometry for the measurement of the three-terminal spin signal. The magnetic field is applied along the z-direction, perpendicular to the spins. (b),(c) Three-terminal Hanle spin signals for the wide (1.2 $\mu$m) contact at 10 K for a current of $-$0.33 mA (b), corresponding to spin extraction from the Si, and for $+$0.33 mA (c), corresponding to spin injection from the Fe into the Si. The pink symbols are the measured spin signal. The green solid lines represent the signal that is expected if the detection efficiency $P_{det}$ is taken to be equal to $P_I$ (linear transport). The blue solid lines represent the signal that is expected if the nonlinear detection efficiency is used. T = 10 K.}
\label{fig7}
\end{figure}

\indent Since all the geometric and spin-transport parameters of our device are known from the nonlocal spin transport data (Fig. 1 and Ref. \onlinecite{spiesserpra}), we exactly know the spin accumulation and its spatial profile, including the magnitude of the spin accumulation under the injector contact. From this, we calculate the magnitude of the three-terminal Hanle signal that is expected from linear transport theory, in which case the detection sensitivity of the contact under bias is equal to the spin polarization of the current at that bias (the latter was presented in Fig. 1d). The expected spin signal is included in Fig. 7 (solid green lines). The result does not match with the data at all. The predicted signal is large for spin injection bias because $P_G$ is large ($P_I$ = $P_{det}$ = 58 \%), so that the induced spin accumulation as well as the spin-detection efficiency are large. However, no spin signal is observed in the experiment. For negative bias the predicted spin signal is rather small because $P_G$ is small for spin extraction ($P_I$ = $P_{det}$ = 24 \%), but the experimentally observed signal is about a factor of 4 larger than predicted.\\
\indent In contrast, if the nonlinearity of spin detection is taken into account, the predicted spin signals (Fig. 7, dark blue lines) are in excellent agreement with the experimental data. For positive bias, even though the spin accumulation is large owing to the large spin polarization of the injected current ($P_I$ = 58 \%), the three-terminal spin signal is expected to be very small because at this bias the spin-detection efficiency is almost zero ($P_{det}$ = 2.9 \%). On the other hand, for negative bias, where $P_I$ = 24 \% and thus the spin accumulation is smaller, the spin signals are large because the detection efficiency is larger than unity ($P_{det}$ = 126 \%). Thus, by taking the spin-detection efficiency into account, the three-terminal spin signals can be described not only qualitatively, but also quantitatively.\\
\indent Two final remarks. Whereas it has been pointed out before that the detected three-terminal spin signal for spin injection bias is unusually small, it has never been pointed out that for spin extraction, the three-terminal spin signal is actually enhanced compared to what is expected from linear transport. Secondly, it is common practise to determine the tunnel spin polarization from three-terminal Hanle signals using linear transport theory, which, given that signals are only observed for spin extraction, will lead to a significant error (i.e., overestimation) of the tunnel spin polarization, and potentially to values of $P_G$ that are larger than 100 \%, which, by definition, is impossible. Our analysis shows that proper values of $P_G$ can be extracted from the three-terminal Hanle data, but only if the spin-detection efficiency is known (from measurements such as those in Fig. 2).

\section{SUMMARY}

It is shown that in a biased ferromagnetic tunnel contact, spin detection is strongly nonlinear. As a result, the spin-detection efficiency is not equal to the tunnel spin polarization. In Si-based 4-terminal spin-transport devices, even a small bias (tens of mV) across the Fe/MgO detector contact enhances the spin-detection efficiency to values up to 140 \% (spin extraction bias) or, for spin injection bias, reduces it to almost zero, while, parenthetically, the charge current remains highly spin polarized. Calculations reveal that the nonlinearity originates from the energy dispersion of the tunnel transmission and the resulting nonuniform energy distribution of the tunnel current, offering a route to engineer spin conversion. The results imply that the description of spin detection in tunnel contacts based on linear transport equations is correct at strictly zero bias and thus can be used to describe conventional nonlocal devices (with an unbiased detector), but it fails to provide a correct description of spin transport in devices with a biased detector, including the two-terminal magnetoresistance. The notion of the inherent nonlinearity of spin detection presents a milestone in our understanding of spin transport in nanodevices and allows a complete, unified and quantitative description of spin transport with consistency between nonlocal spin signals and spin signals in devices in which the detector is biased. Although the results were obtained on devices with a silicon channel, the described phenomena originate from the tunneling process and are therefore present for any type of channel, including other semiconductors, metals, graphene and other two-dimensional materials.

\section{ACKNOWLEDGMENTS}
This work was supported by the Grant-in-Aid for Scientific Research on Innovative Areas, “Nano Spin Conversion Science” (Grants No.26103002 and 26103003).


\begin{thebibliography}{99}
\bibitem{notePdet} Note that the data in Fig. 6e and 6i correspond to a non-local measurement with a biased detector, with the
detector contact biased in spin injection mode, whereas the spin accumulation in the channel is created with the other contact
in spin extraction mode. The latter is opposite to the measurement in Fig 2, in which the spin accumulation is induced with the
injector biased in spin injection mode. The fact that for the data in 6e and 6i the spin signals are absent (i.e., the spin-detection
efficiency is near zero), shows that the spin-detection efficiency does not depend on whether the spin accumulation in the channel
is created by spin injection or spin extraction at the injector.
\bibitem{noteE} We computed the electric field in the lateral direction of the Si channel, because this produces the largest current density. The
electric field perpendicular to the Si channel, at the tunnel interface, is even weaker, because the current is the same but the area of the tunnel contact is more than an order of magnitude larger than the cross section of the Si channel (compare the width of the ferromagnetic contact (1.2 $\mu$m) to the thickness of the Si channel (70 nm)).
\bibitem{note65} It was confirmed that the calculated spin-detection efficiency is directly proportional to $P_G$. We choose to display the results obtained
for a rather large value of $P_G$ (65 \%) in order to illustrate that the calculated spin-detection efficiency can be larger than 100 \%.
\bibitem{noteE330rt} The conclusion that spin drift is negligible can also be obtained by comparing the electric field to the critical field $E_{crit}$ above which
spin drift becomes relevant\cite{flatte}. At room temperature, the electric field in the Si channel is 2.45 kV/m for a current of 0.33 mA (using $\rho_{Si}$ = 2.6 $m\Omega$cm). With $D$ = 4 cm$^2$/s, $\mu_e$ = 150 cm$^2$/Vs and $L_{SD}$ = 1.0 $\mu$m at room temperatue, we obtain $\varepsilon_{drift}$ = 26.7 meV and $E_{crit}$ = 26.7 kV/m. Spin drift is thus negligible at electric fields of 2.45 kV/m (it changes the spin-transport length by less than 5 \%).
\bibitem{notenondegenerate} Although electric fields and spin drift certainly play a role when the channel is a nondegenerate
semiconductor\cite{shiraishipra,shiraishitahara}, the nonlinear spin detection effficiency should also be taken into account for nondegenerate semiconductors and explains the absence of a spin signal at the spin injection contact, irrespective of whether spin drift is present or not.
\bibitem{notehanle} When the lateral dimensions of the ferromagnetic contact are comparable to or smaller than $\lambda_{SD}$, the Hanle curve for a
three-terminal measurement is not exactly Lorentzian and the line width depends on the lateral dimensions of the contact. The width of the Hanle curve was therefore obtained from a full numerical calculation that includes the integration over the spatial variables \cite{spiesserpra}.
\end{thebibliography}
\end{document}